# Ferromagnetism in Dilute Magnetic Semiconductors

R. da Silva Neves[a], A. Ferreira da Silva[a] and R. Kishore[b,*]

[a] Instituto de Física, Universidade Federal da Bahia, Campus Ondina
40210 340 Salvador, Bahia, Brazil

[b] Instituto Nacional de Pesquisas Espaciais – INPE/LAS
12210 970 São José dos Campos, São Paulo, Brazil.

**Abstract**

We study the ferromagnetism of $Ga_{1-x}Mn_xAs$ by using a model Hamiltonian, based on an impurity band and the anti-ferromagnetic exchange interaction between the spins of Mn atoms and the charge carriers in the impurity band. Based on the mean field approach we calculate the spontaneous magnetization as a function of temperature and the ferromagnetic transition temperature as a function of the Mn concentration. The random distribution of Mn impurities is taken into account by Matsubara and Toyozawa theory of impurities in semiconductors. We compare our results with experiments and other theoretical findings.

## 1. Introduction

Dilute magnetic semiconductors (DMS) are promising for technological applications as well as interesting from the basic physics point of view. Possible applications exist in spin electronics (spintronics) [1]. which employ the spin degree of freedom of electrons in addition to their charge. This may allow the incorporation of ferromagnetic elements into semiconductor devices, and thus the integration of data processing and magnetic storage on a single chip. Since the electronic spin is a quantum mechanical degree of freedom, quantum interference effects could be exploited in devices, eventually leading to the design of quantum computers [2]. A lot of work has been done on disorder effects in nonmagnetic semiconductors and metals [3]. Only, during the last few years disorder effects in DMS have been considered. They can be strong due to the presence of a high concentration of charged impurities; the typical distance between these defects is roughly of the same order as the Fermi wavelength.

A key component of spintronics is the development of new ferromagnetic semiconductors. Following the successful development of $Ga_{1-x}Mn_xAs$ [4] and $In_{1-x}Mn_xAs$ [5] as ferromagnetic semiconductors (with x ~ 1- 10%) using careful low temperature molecular beam epitaxy (MBE) technique, intensive worldwide activity has led to claims of ferromagnetism (some at room temperature and above) in several DMS like GaMnP [6], GaMnN [7], GeMn [8], GaMnSb [9] etc. It is at present unclear, whether all these reports of ferromagnetism (particularly at room temperatures or above) are indeed intrinsic magnetic behavior or are arising from clustering and segregation effects associated with various Mn complexes (which have low solubility) and related material problems. The observed ferromagnetism of $Ga_{1-x}Mn_xAs$ is, however, well established and is universally believed to be an intrinsic DMS phenomenon.

The Mn dopants in GaMnAs serve the dual roles of magnetic impurities providing the local magnetic moments and of acceptors producing, in principle, one hole per Mn atom. The number density $n_h$ of the charge carriers (holes in GaMnAs), however, turns out to be by almost an order of magnitude lower than the number density $n_i$ of the Mn ions. The precise role played by the relative values of $n_i$ and $n_h$ in giving rise to DMS ferromagnetism is currently being debated in the literature to the extent that there is no agreement even on whether the low density of charge carriers ( $n_h \ll n_i$ ) in the system helps or hinders ferromagnetism. An important question in this context is to obtain the functional dependence of the ferromagnetic transition temperature $T_C$ and the magnetic moment on hole densities $n_h$ and the Mn ion number density $n_i$.

The existence of DMS ferromagnetism seems to be independent of the system being metallic or insulating. For GaMnAs, both metallic and insulating samples are ferromagnetic with the transition temperature being typically higher for metallic systems although this may not necessarily be a generic behavior. Many other systems, exhibiting DMS ferromagnetism [6-9], are however strongly insulating. Additionally, in $Ga_{1-x}Mn_xAs$ samples, exhibiting reentrant metal-insulator transitions, the ferromagnetic behavior appears to be completely continuous as a function of x with the only observable effect being a variation in $T_C$. The real DMS carriers mediating the ferromagnetic interaction between the local moments are generally likely to be far from free holes in the valence band of the host semiconductor material. They are, in all likelihood, extended or bound carriers in the impurity band, which forms in the presence of the Mn dopants. There is a great deal of direct experimental support for the relevance of this impurity band picture for DMS ferromagnetism, both first principle band structure calculations [10] and Monte Carlo simulation [11] confirm the impurity band nature of the carriers active in DMS ferromagnetism.

---

[*] Corresponding author.
Tel:55-12-32086714; E-mail : rkishore.br@gmail.com

In the wake of Matsubara and Toyozawa method (MT) for a disordered system [12], we study the magnetic properties of GaMnAs, described by states of holes forming an impurity band. Following the mean field approximation, we compare our results with that of Berciu and Bhatt (BB) [13]. There are two main differences between our approach and that of BB [13]: 1. we perform our calculations considering an infinite lattice and 2. we model an effective antiferromagnitic (AFM) exchange interaction coupling between the impurity and holes spins considering the Mn superlattice as an average of SC, BCC and FCC lattices, which is a simple estimate of the random distribution of Mn impurities in GaAs.

In Section 2 we describe the model Hamiltonian, the mean field approximation, and the Matsubara-Toyozawa approach to consider the random distribution of the magnetic impurities. In section 3 we present the calculations and discussions of our results. Finally in Section 4 we give a brief summary of our work.

## 2. Formalism

Following (BB) [13], we use the electron formalism, although the charge carriers in $Ga_{1-x}Mn_xAs$ are holes. It is an equivalent system doped with hypothetical donors, with impurity levels below a conduction band. Since in this work, our main interest is to consider the effect of disorder, this will not change the essential aspects of the problem, as discussed by BB [13]. We consider exactly the same model Hamiltonian as that of BB [13]. In presence of the magnetic field $H$, this Hamiltonian is given as

$$H = \sum_{ij\sigma} t_{ij} c_{i\sigma}^\dagger c_{j\sigma} + \sum_{ij} J_{ij} \mathbf{S}_i \cdot \mathbf{s}_j - g\mu_B H \sum_i s_i^z - g'\mu_B H \sum_i S_i^z \quad (1)$$

where $c_{i\sigma}^\dagger$ and $c_{i\sigma}$ are the creation and annihilation operators for holes of spin $\sigma$ at the site I, and $J_{ij}$ is the anti-ferromagnetic (AFM) exchange interaction between spins of holes and Mn ions. The first term describes the charge carriers hopping between Mn sites with a hopping $t_{ij}$, dependent on the distance $|\mathbf{R}_i - \mathbf{R}_j|$. We restrict ourselves to the uncompensated samples in which $t_{ii}$ can be considered a constant independent of site $i$. We shall define the energy scale such that $t_{ii} = 0$. The transfer integral $t_{ij}$ will be calculated from 1s hydrogen like wave functions. The second term describes the AFM interaction between $\mathbf{S}_{Mn}$ and $\mathbf{s}_h$ spins, and the last two terms describe the energy interaction with the external magnetic field.

We treat the second term in the molecular field approximation as

$$\mathbf{S}_i \cdot \mathbf{s}_j = S_{Mn} s_j^z + s_h S_i^z - S_{Mn} s_h \quad (2)$$

where $S_{Mn} = <S_i^z>$ and $s_h = <s_j^z>$ are assumed to be configurationally averaged values of Mn and hole spins and are thus independent of the site index. Within this mean field approximation, the Hamiltonian (1) becomes

$$\mathbf{H} = \sum_i \left[ \left(J_{eff} s_h - g'\mu_B H\right) \mathbf{S}_i^z - g\mu_B H \mathbf{s}_i^z \right] + \sum_{i\sigma} \frac{\sigma}{2} J_{eff} S_{Mn} c_{i\sigma}^\dagger c_{i\sigma} + \sum_{ij\sigma} t_{ij} c_{i\sigma}^\dagger c_{j\sigma} - J_{eff} S_{Mn} \quad (3)$$

where $J_{eff} = \sum_i J_{ij} = J \sum_i exp(-2|\mathbf{R}_i - \mathbf{R}_j|/a_B)$.

and $a_B$ is the Bohr radius. For this molecular field Hamiltonian, we get

$$S_{Mn} = B_S [\beta(g'\mu_B H - s_h J_{eff})], \quad (4)$$

where, $B_S(x) = (S+1/2) coth[(S+1/2)x] - (1/2) coth(x/2)$ is the Brillouin function and $\beta = 1/(k_B T)$. Following BB [13] we have assumed that for Mn atoms spin S = 5/2.

Now to obtain $s_h$ we consider the Green's function [14]

$$G_{ij\sigma}(t) = -i\theta(t) < [c_{i\sigma}(t), c_{j\sigma}^\dagger]_+ > \quad (5)$$

where, $[A, B]_\sigma = AB + \sigma BA$; $\sigma$ = + or -. By considering the Fourier transform

$$G_{ij\sigma}(t) = (1/2\pi) \int_{-\infty}^{\infty} G_{ij\sigma}(\omega) e^{-i\omega t} dt \quad (6)$$

we get the equation of motion of the Green's function $G_{ij}(\omega)$ as

$$(\omega - t_\sigma) G_{ij\sigma}(\omega) = \delta_{ij} + \sum_l t_{il} G_{lj\sigma}(\omega) \quad (7)$$

where

$$t_\sigma = (\sigma/2)(S_{Mn} J_{eff} - g\mu_B H). \quad (8)$$

Now, considering the completely random distribution of the magnetic impurities, we use the approach of Matsubara and Toyozawa to obtain the configurationally averaged diagonal Green`s function as [12,14]

$$<G_{ii\sigma}(\omega)> = (1/(\omega - t_\sigma))\, \zeta(\omega - t_\sigma) \quad (9)$$

where

$$\zeta(\omega) = 1/(1 - \eta(\omega)) \quad (10)$$

and

$$\eta(\omega) = \frac{n_i \zeta(\omega)}{8\pi^3 \omega^2} \int dk \, \frac{t^2(\mathbf{k})}{1 - \frac{n_i \zeta(\omega)}{\omega} t(\mathbf{k})} \quad (11)$$

where $t(\mathbf{k})$ is the Fourier transform of

$$t_{ij} = -2t_0 (1 + a_B^{-1} |\mathbf{R}_i - \mathbf{R}_j|) \exp(-a_B^{-1}|\mathbf{R}_i - \mathbf{R}_j|) \quad (12)$$

in the reciprocal space. Here $t_0$ is the ionization energy of 1s state.

The averaged Green function $<G_{ii\sigma}(\omega)>$ can be used to obtain the density of states per impurity for spin $\sigma$ as

$$D_\sigma(\omega) = -(1/\pi)\lim_{\varepsilon \to 0} \mathrm{Im}<G_{ii\sigma}(\omega + i\varepsilon)> \quad (13)$$

We use this density of states to obtain $s_h$ as

$$s_h = (1/2)\sum_\sigma \sigma \int d\omega\, D_\sigma(\omega)\, (e^{\beta(\omega-\mu)} + 1)^{-1} \quad (14)$$

where the Fermi energy $\mu$ can be obtained from the number of holes per impurity $p = n_h / n_i$ as

$$p = \sum_\sigma \int d\omega\, D_\sigma(\omega)\, (e^{\beta(\omega-\mu)} + 1)^{-1} \quad (15)$$

## 3. Results and Discussions

For our calculations, we take the same parameters as used by BB [13]. For example we take, the ionization energy $t_0 = 112.4$ meV=1Ry, the Bohr radius $a_B = 7.8$ Å and J = 15 meV. The number density $n_i$ of the Mn impurities is calculated from the expression $n_i = 4x/a^3$, where $a = 5.65$ Å, is the lattice constant of GaAS lattice. Now, first we fix the value of $x$ to calculate $n_i$ and assume the initial value of $S_{Mn} = 5/2$. This enable us to calculate $D_\sigma(\omega)$ and $\mu$ from Eqs. (13) and (15) respectively and then $s_h$ from Eq. (14) for a fixed value of temperature T. The value of $s_h$ obtained is used to calculate $S_{Mn}$ from Eq. (4). This procedure is repeated till self-consistency is achieved. This allowed us to calculate $S_{Mn}$ and $s_h$ as a function of $k_B T/J$ for various value of $p$ and $x$ and compare it with the results of BB [13].

We have also done our self consistent calculations, considering that Mn impurities are placed in an ordered lattice. We assume that this ordered lattice is an average of SC, BCC and FCC lattices in contrast to BB [13] who considered that the Mn impurities are distributed in SC structure. For ordered system, Eqs. (14) and (15) reduce to

$$\mathbf{s}_h = (16\pi^3)^{-1} \int d\mathbf{k}\, \sigma \sum_\sigma f(t_{\mathbf{k},\sigma}) \quad (16)$$

$$p = (8\pi^3)^{-1} \int d\mathbf{k} \sum_\sigma f(t_{\mathbf{k},\sigma}) \quad (17)$$

where $t_{\mathbf{k}\sigma} = t_\mathbf{k} + t_\sigma$ and $f(t_{\mathbf{k}\sigma}) = (1 + e^{\beta(t_{\mathbf{k}\sigma}-\mu)})^{-1}$ is the Fermi distribution function. The expression for $S_{Mn}$ is given by Eq. (4).

We have shown our results in Figures 1-4. In Fig. 1, we have plotted the density of states (DOS) for a completely random distribution of Mn impurities corresponding to x = 0.05 and p = 0.1 and compared it with that of BB [13]. In Fig 2 our results show that the disordered system has a significant difference in the behavior of magnetization curves compared to the ordered case. According to BB [13] and C. Timm [1], in disordered systems the average magnetization of Mn spins as a function of temperature has more pronounced decrease in the low temperatures and decreased more slowly in the region of intermediate and high temperatures, leading to higher ferromagnetic transition temperature $T_C$ [15] than in the case of ordered systems. Although $T_C$ present higher values in disordered systems, the magnetization due to Mn spins falls dramatically from the saturation value, unlike ordered case where there is a less marked fall in the region of low temperatures with a considerable magnetization for intermediate temperatures.

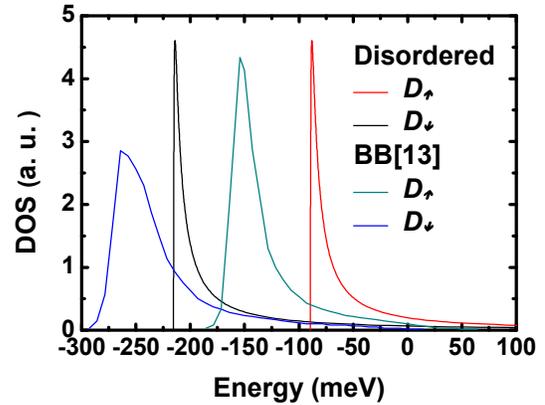

FIG. 1 Impurity band density of states for the spin splitting at $T = 0K$ with $x = 0.05$ and $p = 0.1$. The calculations from BB [13] for the completely random distribution of impurities are also presented.

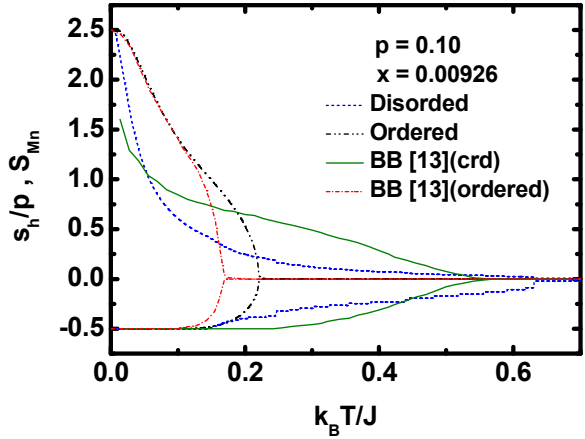

FIG. 2 The average Mn and carrier spins $S_{Mn}$ and $s_h/p$ for both ordered and disordered systems as a function of $k_B T/J$. For the sake of comparison the results of BB [13] for both ordered and disordered system with completely random distribution (crd) of impurities are also shown.

Fig.3 shows the effect of magnetic field on $S_{Mn}$ and $s_h/p$. It shows that the application of magnetic field influences more strongly the spins of Mn impurities than hole spins. It should be noted that, except at very low temperatures and very near to transition temperature, our values for $S_{Mn}$ ($s_h/p$) are always lower (higher) than that of BB [13].

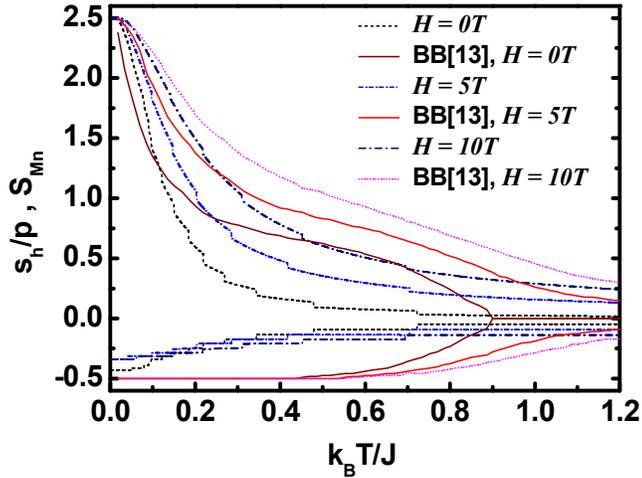

FIG. 3 The average Mn and carrier spins $S_{Mn}$ and $s_h/p$ as a function of $k_B T/J$ with and without application of an external magnetic field. For comparison, the results of BB [13] are also shown.

In Fig. 4 we have shown calculation of the transition temperature $T_c$ as a function of x for both ordered and disordered samples. Results are compared with the experimental results of Matsukura et al. [16] and Beschoten et al. [17]. Although $T_C$ has to be higher for the disordered system, the spontaneous magnetization has a residual value on a large extension of the temperature values. This can be explained by the formation of clusters of impurities. Each cluster can have a magnetization in a given direction, providing several different regions with magnetization in random directions, which can be explained by the long tail found in the curves of spontaneous magnetization for disordered systems.

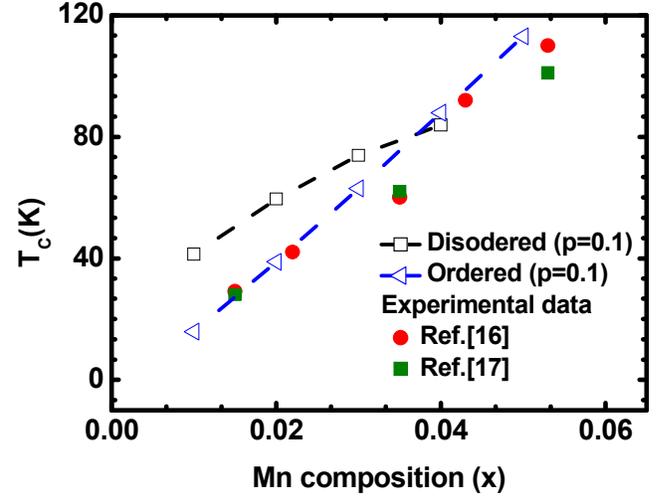

FIG. 4 Comparison between theoretical (disordered and ordered) and the experimental values of $T_C$.

## 4. Conclusions

Briefly we have studied the disorder effects in GaMnAs dilute magnetic semiconductors described by an impurity band using Matsubara and Toyozawa approach for disordered semiconductors. We have compared our results with that of BB [13] and the experimental results of $T_c$. Since this model is applicable for the infinite size of the sample, it can be useful to clarify some points that are not yet well understood because of the finite size calculations.

## Acknowledgements

This work was partially supported by the Brazilian agencies FAPESB and CNPQ.